\documentclass[%
 reprint,
 amsmath,amssymb,
 aps,
]{revtex4-2}

\usepackage{graphicx}
\usepackage{dcolumn}
\usepackage{bm}
\usepackage{color}
\usepackage{cancel}
\usepackage{comment}

\newcommand{\del}{\partial}
\newcommand{\beq}{\begin{eqnarray}}
\newcommand{\eeq}{\end{eqnarray}}

\newcommand{\tr}{\mathop{\mathrm{tr}}}
\newcommand{\SU}{\text{SU}}
\newcommand{\U}{\text{U}}
\newcommand{\rmi}{\text{i}}
\newcommand{\rme}{\text{e}}
\newcommand{\rmd}{\text{d}}

\begin{document}

\preprint{
YGHP-23-01, KEK-TH-2513
}

\title{How baryons appear in low-energy QCD:\\
Domain-wall Skyrmion phase in strong magnetic fields 
}

\author{Minoru Eto$^{1,2}$, Kentaro Nishimura$^{3,2}$, Muneto Nitta$^{2,4,5}$}

\affiliation{%
$^1$Department of Physics, Yamagata University, 
Kojirakawa-machi 1-4-12, Yamagata,
Yamagata 990-8560, Japan, \\
$^2$Research and Education Center for Natural Sciences, Keio University, 4-1-1 Hiyoshi, Yokohama, Kanagawa 223-8521, Japan\\
$^3$KEK Theory Center, Tsukuba 305-0801, Japan\\
$^4$Department of Physics, Keio University, 4-1-1 Hiyoshi, Yokohama, Kanagawa 223-8521, Japan\\
$^5$
International Institute for Sustainability with Knotted Chiral Meta Matter(SKCM$^2$), Hiroshima University, 1-3-2 Kagamiyama, Higashi-Hiroshima, Hiroshima 739-8511, Japan
}%

\date{\today}

\begin{abstract}
Low-energy dynamics of QCD can be described 
by pion degrees of freedom in terms of  
the chiral perturbation theory(ChPT).
A chiral soliton lattice(CSL), an array of solitons, is the ground state 
due to the chiral anomaly 
in the presence of a magnetic field 
larger than a certain critical value 
at finite density.
Here, we show 
in a model-independent and fully analytic manner (at the leading order of ChPT) that the CSL phase transits to
a {\it domain-wall Skyrmion phase} 
when the chemical potential is 
larger than the critical value
$\mu_{\rm c} = 16\pi f_{\pi}^2/3m_{\pi}
    \sim 1.03 \;\; {\rm GeV}$ 
    with the pion's decay constant 
    $f_{\pi}$ and mass $m_{\pi}$, 
    which can be regarded as the nuclear saturation density.
There spontaneously appear
stable two-dimensional Skyrmions or lumps 
 on a soliton surface,  
which can be viewed 
as three-dimensional Skyrmions 
carrying even baryon numbers 
from the bulk despite no Skyrme term. 
They behave as superconducting rings 
with persistent currents due to a charged pion condensation, 
and areas of the rings' interiors are quantized. 
This phase is in scope of future heavy-ion collider experiments.
\end{abstract}

\maketitle


\section{Introduction}

We are all made of baryons, 
that is, nucleons such as protons and neutrons, 
and nucleons are composed of quarks and gluons, 
particles gluing quarks.
Theoretically, these particles are 
described by
Quantum Chromodynamics (QCD), 
fundamental theory of the strong interaction. 
It is, however, 
 quite difficult to prove by first-principles calculation 
 of QCD that 
 quarks are all confined to form 
hadrons, {\it i.~e.}
baryons (three quark bound states 
such as nucleons) 
and mesons (quark--anti-quark bound states such as pions).
Nevertheless, low-energy dynamics of QCD 
 can be described in terms of 
 symmetry:
chiral symmetry, 
a symmetry mixing different species of 
quarks (up-quarks, down-quarks and so on)
 in QCD, 
is spontaneously broken with   
resulting in light scalar 
(Nambu-Goldstone) bosons, that is 
pions. 
Thus, low-energy dynamics of QCD can be described 
by these pion degrees of freedom  
in a model-independent manner in terms of 
a chiral Lagrangian or 
more generally within the chiral perturbation theory (ChPT) 
up to some constants, the pion's decay constant, 
quark masses and so on 
\cite{Scherer:2012xha,Bogner:2009bt}.

However, 
one of big questions is 
how baryons 
can be described at low energy.
One of old ideas by Skyrme was 
that nucleons can be identified with 
topological solitons, 
called Skyrmions, 
of the chiral Lagrangian
supplemented by the so-called four derivative 
Skyrme term stabilizing Skyrmions 
\cite{Skyrme:1962vh,Witten:1983tx}.
However, one of the drawbacks of the Skyrme model 
is the necessity of the Skyrme term, 
a particular choice of four derivative term;
all quantities depend on 
the coupling constant of 
the Skyrme term, 
and thus all predictions are 
model-dependent in this sense.

One of breakthroughs to overcome the problem was made by 
Son and Stephanov \cite{Son:2007ny}.
The chiral Lagrangian 
contains the anomalous coupling of the neutral pion $\pi_0$ to the magnetic field via the chiral anomaly 
\cite{Son:2004tq,Son:2007ny}
in terms of   
the Goldstone-Wilczek current \cite{Goldstone:1981kk,Witten:1983tw}.
It has been pointed out that, due to this anomalous term, the ground state of QCD with two flavors (up and down quarks) at a finite baryon chemical potential $\mu_{\textrm{B}}$ under a sufficiently strong magnetic field is a stack of sheet-type baryons, which is called chiral soliton lattice (CSL).
Focusing only on the neutral pion $\pi_0$,
the ChPT is mathematically equivalent to the sine-Gordon model and has the domain wall solution corresponding to the sheet-type baryons \cite{Son:2007ny,Eto:2012qd,Brauner:2016pko}. 
Recently, CSL phases have been paid great interests: 
CSLs appear also  
under thermal fluctuation \cite{Brauner:2017uiu,Brauner:2017mui,Brauner:2021sci,Brauner:2023ort}
or rapid rotation 
\cite{Huang:2017pqe,Nishimura:2020odq,Eto:2021gyy,Chen:2021aiq}.
Other topics include 
the instability of CSLs via a charged pion condensation \cite{Brauner:2017uiu}, 
a possibility of an Abrikosov's vortex lattice \cite{Evans:2022hwr}, 
relations between Skyrmions and CSLs \cite{Kawaguchi:2018fpi,Chen:2021vou,Chen:2023jbq},
and quantum nucleation of CSLs 
\cite{Eto:2022lhu,Higaki:2022gnw}
(see also \cite{Yamada:2021jhy,Brauner:2019aid,Brauner:2019rjg}).
Furthermore, apart from QCD, CSLs
universally appear in various condensed matter systems; 
chiral magnets 
with nanotechnological application 
to magnetic memory storage devices and magnetic sensors \cite{togawa2016symmetry},
and chiral liquid crystals.

In this Letter, we find 
a new baryonic phase   
 in a model-independent and analytic manner, 
that is a 
{\it domain-wall Skyrmion phase} 
appearing inside higher density region of the CSL
with the baryon chemical potential 
\begin{gather}
    \mu_{\rm B} \geq \mu_{\textrm{c}} = \frac{16\pi f_{\pi}^2}{3m_{\pi}}
    \sim 1.03 \;\; {\rm GeV} \,, \label{eq:negative}
\end{gather}
where we have used the vacuum values of the physical quantities $f_{\pi}\approx 93\, \textrm{MeV}$ and $m_{\pi}\approx 140\, \textrm{MeV}$.
This phase boundary coincides with 
the instability of CSLs via a charged pion condensation 
\cite{Brauner:2017uiu}. 
In this phase,
two-dimensional Skyrmions \cite{Polyakov:1975yp}
spontaneously appear on a soliton surface, 
which can be viewed 
as three-dimensional Skyrmions 
from the bulk, 
thereby called domain-wall Skyrmions 
\cite{Nitta:2012wi,Nitta:2012rq,
Gudnason:2014nba,Gudnason:2014hsa}.

\section{Chiral solitons lattices}
We focus on the phase in which the chiral symmetry is spontaneously broken down.
The low-energy dynamics can thus be described by an effective field theory of the pions -- ChPT.
The pion fields are represented by a $2\times 2$ unitary matrix
\begin{gather}
    \Sigma = \rme^{\rmi \tau_a\chi_a} \,,
\end{gather}
where $\tau_a$ ($a=1,2,3$) are the Pauli matrices with the normalization $\tr(\tau_a\tau_b)=2\delta_{ab}$.
This field $\Sigma$ transforms under $\SU(2)_{\textrm{L}}\times \SU(2)_{\textrm{R}}$ chiral symmetry as
    $\Sigma \to L\Sigma R^{\dag}$,
where $L$ and $R$ are $2\times 2$ unitary matrices.
Then, the effective Lagrangian 
at the ${\cal O}(p^2)$ order is 
($\mu=0,\cdots,3$)
\begin{gather}
    \mathcal{L}_{\textrm{ChPT}}
    = \frac{f_{\pi}^2}{4} \tr \left(D_{\mu}\Sigma D^{\mu}\Sigma^{\dag} \right) 
    - \frac{f_{\pi}^2m_{\pi}^2}{4}(2-\Sigma-\Sigma^{\dag}) \label{ChPT_with_B} \,,
\end{gather}
where $f_{\pi}$ and $m_{\pi}$ are pion's decay constant and mass, respectively, and 
$D_{\mu}$ is a covariant derivative defined by
\begin{gather}
    D_{\mu}\Sigma \equiv \del_{\mu}\Sigma + \rmi e A_{\mu} [Q, \Sigma] \label{def_cov_del} \,,\quad
    Q 
    = \frac{1}{6}\bm{1} + \frac{1}{2}\tau_3 \,.
\end{gather}
The $U(1)_{\rm EM}$ transformation is
$\Sigma\to \rme^{i\lambda\frac{\tau_3}{2}} \Sigma \rme^{-i\lambda\frac{\tau_3}{2}}$ and $A_\mu \to A_\mu - \frac{1}{e}\partial_\mu \lambda$.
The external $\U(1)_{\textrm{B}}$ gauge field $A^{\textrm{B}}_{\mu}$ can couple to  $\Sigma$ through the Goldstone-Wilczek (GW) current \cite{Goldstone:1981kk,Witten:1983tw}.
The conserved and gauge-invariant baryon current in the external magnetic field is calculated in refs.~\cite{Goldstone:1981kk,Son:2007ny}:
\begin{gather}
\hspace{-.3cm}
    j_{\textrm{GW}}^{\mu}
    = -\frac{\epsilon^{\mu \nu \alpha \beta}}{24\pi^2} \tr \{
    L_{\nu}L_{\alpha}L_{\beta}
    - 3\rmi e \del_{\nu} \left[A_{\alpha} Q(L_{\beta} + R_{\beta}) \right]
    \} \hspace{-.3cm}
\label{eq:GW}
\end{gather}
where $A^{\textrm{B}}_{\mu}=(\mu_{\textrm{B}}, \bm{0})$,
and we have introduced the standard notation $L_{\mu}\equiv \Sigma \del_{\mu}\Sigma^{\dag}$ and $R_{\mu}\equiv \del_{\mu}\Sigma^{\dag}\Sigma$.
Then, the effective Lagrangian coupling to $A^{\textrm{B}}_{\mu}$ can be written as
\begin{gather}
    \mathcal{L}_{\textrm{WZW}} = -A^{\textrm{B}}_{\mu}j_{\textrm{GW}}^{\mu} \label{effective_Lagrangian_GW} \,.
\end{gather}
which is known as 
the Wess-Zumino-Witten(WZW) term
\cite{Son:2004tq,Son:2007ny}.
The total Lagrangian is
$\mathcal{L}
    \equiv \mathcal{L}_{\textrm{ChPT}} + \mathcal{L}_{\textrm{WZW}}$.

An important remark is in order here.
To construct an effective Lagrangian,
we choose a modification of the standard power counting scheme of ChPT \cite{Brauner:2021sci} :
\begin{gather}
    \del_{\mu} \,, m_{\pi} \,, A_{\mu} = \mathcal{O}(p^1) \,, \qquad
    A^{\textrm{B}}_{\mu} = \mathcal{O}(p^{-1}) \,.
\end{gather}
We note that, in this power counting, eq.~(\ref{effective_Lagrangian_GW}) has the order $\mathcal{O}(p^2)$ and is of the same order as eq.~(\ref{ChPT_with_B}).
The fact that $\mu_{\textrm{B}}$ appears only in the WZW term in eq.~(\ref{effective_Lagrangian_GW}) enables us to assign the negative power counting to $\mu_{\rm B}$.
The effective field theory up to $\mathcal{O}(p^2)$ has to contain both terms in eq.~(\ref{eq:GW}); however, the first term in eq.~(\ref{eq:GW}) has not been considered in the previous researches of the CSLs. 
We emphasize that we do not need 
an $\mathcal{O}(p^4)$ order such as a Skyrme term to 
obtain our results, 
and thus our analysis is model-independent.

We note that our effective theory admits the sine-Gordon soliton solutions,
and they are stable under a sufficiently large magnetic field as shown in \cite{Son:2007ny}.
If we consider the case of 
no charged pions 
$\Sigma = \rme^{\rmi \tau_3\chi_3}$, the effective Lagrangian reduces to 
\begin{eqnarray}
\mathcal{L}
= \frac{f_{\pi}^2}{2}(\del_{\mu}\chi_3)^2
    - f_{\pi}^2m_{\pi}^2(1-\textrm{cos}\chi_3)
    + \frac{e\mu_{\textrm{B}}}{4\pi^2}\bm{B}\cdot \bm{\nabla}\chi_3 \,.
\end{eqnarray}
The sine-Gordon Lagrangian consists of 
the first and second terms, while 
the third term is a topological term 
that does not contribute to the EOM. 
Nevertheless 
it stabilizes an inhomogeneous configuration of $\chi_3$.
Solutions of $\chi_3$ are sine-Gordon solitons winding around $\U(1)_3 \subset \SU(2)_{\textrm{V}}$.
A single soliton solution is 
\begin{gather}
    \chi_3 = 4\tan^{-1}e^{m_{\pi}(z-Z)} \label{SG_soliton} \,,
\end{gather}
where $Z$ is the position of the center 
(or the translational modulus).
The tension, that is 
the energy density per unit area, can be analytically obtained:
\begin{align}
    E 
    = 8m_{\pi}f_{\pi}^2 - \frac{e\mu_{\textrm{B}}B}{2\pi} \label{energy_of_DW} \,.
\end{align}
When $E<0$,
the sine-Gordon soliton is energetically more stable than the QCD vacuum ($\chi_3$=0).
The critical magnetic field at which the transition happens is 
\begin{gather}
    B_{\textrm{c}} = \frac{16\pi m_{\pi}f_{\pi}^2}{e\mu_{\textrm{B}}} \label{B_CSL} \,.
\end{gather}
The ground state at $B>B_{\textrm{c}}$ is a stack of parallel $\chi_3$ solitons,
which is the so-called CSL \cite{Son:2007ny,Eto:2012qd,Brauner:2016pko}.

\section{Non-Abelian Sine-Gordon soliton 
and its effective field theory
}
So far, we 
have neglected the charged pions, $\Sigma=\rme^{i\tau_3\chi_3}$.
General solutions containing charged pions  can be obtained 
from $\Sigma_0$
by an $\SU(2)_{\textrm{V}}$ transformation,
\begin{gather}
    \Sigma = g\Sigma_0 g^{\dag} = \exp(\rmi \theta g \tau_3 g^{\dag}) \label{general_sol} \,,
\end{gather}
where $g$ is an $\SU(2)$ matrix.
Since $g$ in eq.~(\ref{general_sol}) 
is redundant with respect to 
a $\U(1)$ subgroup generated by $\tau_3$,
it takes a value in a coset space, $\SU(2)_{\textrm{V}}/\U(1)\simeq \mathbb{C}P^1 
\simeq S^2$.
Together with the translational modulus $Z$, 
the single  sine-Gordon soliton has the moduli 
\begin{gather}
    \mathcal{M} \simeq \mathbb{R} \times \mathbb{C}P^1 \,.
\end{gather}
Such a soliton with non-Abelian moduli 
is called a non-Abelian sine-Gordon soliton 
\cite{Nitta:2014rxa,Eto:2015uqa}.
For later convenience, we parameterize the $\mathbb{C}P^1$ moduli 
by the homogeneous coordinates 
$\phi \in \mathbb{C}^2$
of $\mathbb{C}P^1$, 
satisfying \cite{Eto:2015uqa}
\begin{gather}
    \phi^{\dag}\phi = 1 \,, \quad\quad
    g \tau_3 g^{\dag}
    = 2 \phi \phi^{\dag} - \bm{1}_2 \,.
\end{gather}
In terms of $\phi$,
eq.~(\ref{general_sol}) can be rewritten as
\begin{gather}
    \Sigma = \exp(2\rmi \theta \phi \phi^{\dag})u^{-1} = [{\bm 1}_2 + (u^2-1)\phi \phi^{\dag}]u^{-1} \label{general_sol_Sigma} \,,
\end{gather}
where we have defined 
$u \equiv \rme^{\rmi \chi_3}
= \rme^{4 {\rmi} \tan^{-1}e^{m_{\pi}(z-Z)}}
$.
We also use the real three-component vector ${\bm n}$ with  $|{\bm n}|=1$ defined by ${\bm n}\cdot{\bm \tau}= 2 \phi\phi^\dag - {\bm 1}_2 = g\tau_3g^\dag$ or $n_a =  \phi^\dag \tau_a\phi$. The single $\pi^0$ soliton (\ref{SG_soliton}) corresponds to $n_3 = 1$.

Now, 
we are ready to construct the low-energy effective theory(EFT) of a single soliton
by using the moduli approximation \cite{Manton:1981mp,Eto:2006uw}.
Let us place a single sine-Gordon soliton perpendicular to the $z$-coordinate.
In the following, we will promote the moduli parameter $\phi$ to be the fields on the $2+1$-dimensional soliton's world volume as
We do not do so for 
the translational modulus $Z$, 
since the transverse motion is 
irrelevant in our study.
By substituting eq.~(\ref{general_sol_Sigma}) into 
${\cal L}$ and integrating over $z$,
we get \footnote{
See Appendix \ref{app:EFT} for 
a derivation of the domain-wall EFT.
} 
\begin{eqnarray}
&&  \mathcal{L}_{\textrm {DW}} = 
        - 8m_{\pi}f_{\pi}^2 + \frac{e\mu_{\textrm{B}}B}{2\pi} +
     \mathcal{L}_{\textrm{norm}} 
     + \mathcal{L}_{\textrm{WZW}} 
     \label{EFT_lag}\\
&&    \mathcal{L}_{\textrm{norm}}
    = 
\frac{16f_{\pi}^2}{3m_{\pi}} [(\phi^{\dag}D_{\alpha}\phi)^2+D_{\alpha}\phi^{\dag}D^{\alpha}\phi]  
   \label{EFT_of_kinetic} \,,
    \quad \\
&&    \mathcal{L}_{\textrm{WZW}}
    = 2\mu_{\textrm{B}}q
    +\frac{e \mu_{\textrm{B}}}{2\pi} \epsilon^{03jk}\del_j[A_k(1-n_3)]  
    \label{EFT_of_GW} \quad \quad
\end{eqnarray}
where $q$ is the lump (baby Skyrmion) 
topological charge density 
for $\pi_2({\mathbb C}P^1)$, defined by
\begin{eqnarray}
    q \equiv
     -\frac{\rmi}{2\pi}\epsilon^{ij}\del_i\phi^{\dag}\del_j\phi
     = \frac{1}{8\pi} \epsilon^{ij}{\bm n}\cdot (\partial_i {\bm n}\times\partial_j{\bm n})
    \label{eq:pi2}
\end{eqnarray}
which arises via the 
baryon number density ${\cal B}$ in the GW current~\footnote{
See Appendix \ref{app:baryon} for a derivation.}
\begin{eqnarray}
    q = \frac{1}{2}\int_{-\infty}^{\infty} \rmd z~{\cal B}     
    \,,\quad
    {\cal B} \equiv \frac{-1}{24\pi^2}\epsilon^{ijk} \tr(L_{i}L_{j}L_{k})\,.
    \label{lump_and_Skyrme}
\end{eqnarray}
In eq.~(\ref{EFT_lag}) the first two terms are constants corresponding to the domain wall energy density in eq.~(\ref{energy_of_DW}).
Eq.~(\ref{EFT_of_kinetic}) corresponds 
to the kinetic terms of $Z$ and $\phi$, respectively, 
in which 
the $U(1)_{\rm EM}$ gauge transformation is given by $\phi \to \rme^{\rmi\frac{\lambda \tau_3}{2}}\phi$ and
$D_{\alpha}\phi= (\partial_{\alpha} + \rmi \frac{e}{2}\tau_3 A_{\alpha})\phi$
is a covariant derivative 
with respect to a background gauge field 
$A_{\alpha}$ in the bulk. 
In eq.~(\ref{EFT_of_GW}),  
the first term 
counts the lump number and
the second is a total derivative term.
Thus, apart from the translational mode, 
the effective theory of $\phi$ is a gauged  ${\mathbb C}P^1$ model 
[or $O(3)$ nonlinear sigma model] with the topological term (\ref{eq:pi2}).

The previous studies 
\cite{Son:2007ny,Eto:2012qd,Brauner:2016pko} 
correspond to 
$n_3=1$ leading 
eq.~(\ref{EFT_of_GW})
to vanish.

\section{Domain-wall Skyrmion phase}
We investigate Skyrmions in 
the domain-wall effective theory 
in eq.~(\ref{EFT_lag}).
To this end, first we turn off the gauge coupling for a while  
$D_{\mu} \to \del_{\mu}$
with taking into account the effects of 
the WZW term in eq.~(\ref{EFT_of_GW}), 
\footnote{
Exactly speaking this corresponds to 
a parameter region that $e\to 0$ with keeping $e\mu$.
}
and later we take into account effects of 
the gauge coupling.
The static Hamiltonian becomes
\begin{eqnarray}
     \mathcal{H}_{\rm DW}
    = \frac{4f_{\pi}^2}{3m_{\pi}}(\del_{i}{\bm n})^2
    -2\mu_{\textrm{B}}q 
    -\frac{e \mu_{\textrm{B}}}{2\pi} \epsilon^{03jk}\del_j[A_k(1-n_3)].
    \label{effective_hamiltonian_without_D} \,
\end{eqnarray}
Since the constants in eq.~(\ref{EFT_lag})
only give the condition of whether the domain wall appears or not,
it is sufficient to consider $B>B_{\rm c}$, 
and thus have been omitted in 
eq.~(\ref{effective_hamiltonian_without_D}). 
Then, the total energy $E_{\rm DW} = \int d^2x\, {\cal H}_{\rm DW}$
is bounded from below as
\begin{eqnarray}
E_{\rm DW} 
 \ge \frac{32f_{\pi}^2\pi |k|}{3m_{\pi}}
    - 2\mu_{\textrm{B}} k   
+ \frac{
    e \mu_{\textrm{B}}B}{4\pi}\oint \rmd S_i\,
    x^i (n_3-1)
    \,,
   \label{Bogomolny_bound}
\end{eqnarray}
where we have used 
\begin{eqnarray}
    \partial_i{\bm n}\cdot \partial_i {\bm n}
    = \frac{1}{2}\left(\partial_i{\bm n} \pm \epsilon_{ij}{\bm n}\times\partial_j{\bm n}\right)^2 \pm
    8 \pi q\,,
\end{eqnarray}
and defined the lump number
\begin{eqnarray}
    k = \int \rmd^2x\, q \in \mathbb{Z}\,.
\end{eqnarray}
The inequality in eq.~(\ref{Bogomolny_bound}) 
is saturated only when the fields satisfy the (anti-)Bogomol'nyi-Prasad-Sommerfield (BPS) equation
\begin{gather}
    \del_i\bm{n}\pm \epsilon_{ij}\bm{n}\times \del_j\bm{n}=0 \,,
\end{gather}
where the upper (lower) sign corresponds to the (anti-)BPS 
equation.
It is interesting to observe that 
the second term in eq.~(\ref{Bogomolny_bound})
splits energies between BPS lumps $k>0$ and 
anti-BPS lumps $k<0$~\footnote{
This situation is similar to magnetic skyrmions 
in chiral magnets, in which case the DM interaction plays 
such a role.
}. 
The BPS solutions to this equation
characterized by the winding number $k$ ($>0$)
is given by \cite{Polyakov:1975yp}
\begin{eqnarray}
        n_3 = \frac{1-|f|^2}{1 + |f|^2}, \quad
        f = \frac{b_{k-1}w^{k-1}+\cdots+b_0}{w^k + a_{k-1}w^{k-1}+\cdots+a_0}\,,
        \label{sol_n}
\end{eqnarray}
where $w \equiv x+\rmi y$, and the set of complex parameters $\{a_A,b_A\}$ ($A=0,1,\cdots, k-1$)
are the moduli parameters.

\begin{figure}
    \centering
    \includegraphics[width=8cm]{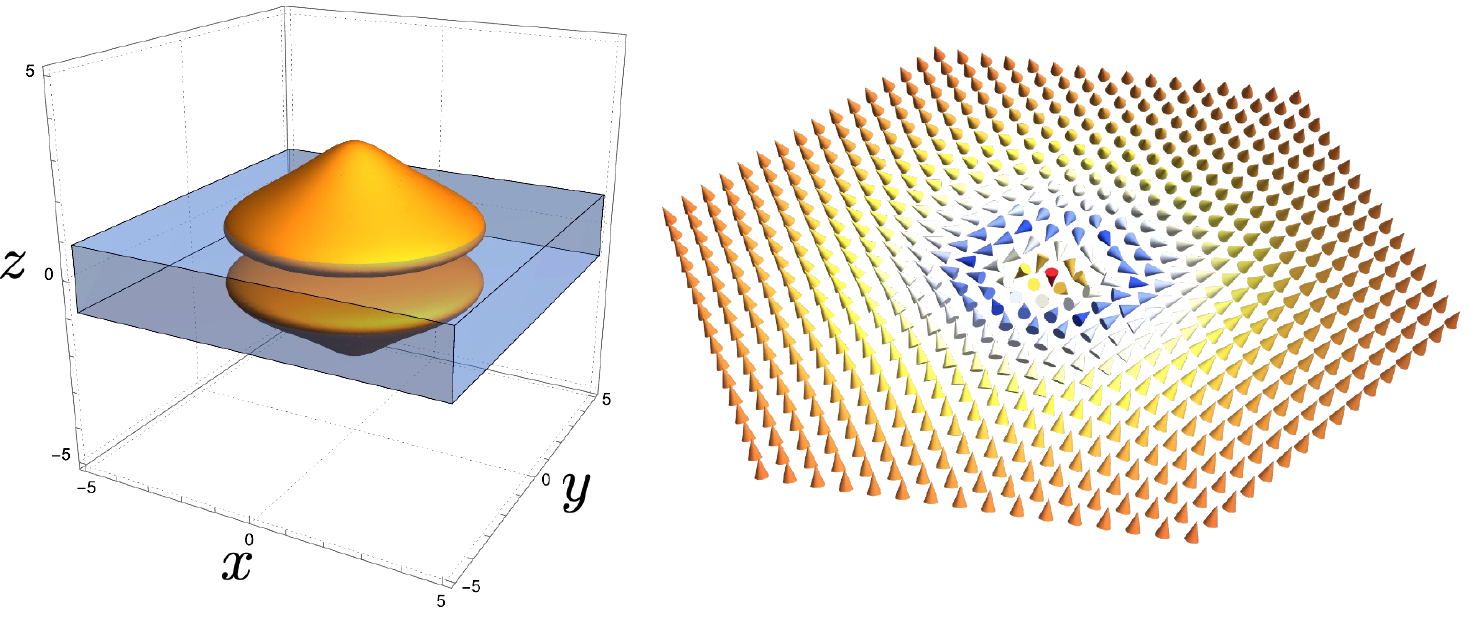}
    \caption{(left) The minimal baryon $f(w) = 1/w$. The isosurface of ${\cal B} = 1/(50\pi^2)$ (orange), and the sine-Gordon soliton $\pi/2 < \theta < 3\pi/2$ (blue). The spatial coordinates are dimensionless with the unit of $m_\pi^{-1}$. (right) ${\bm n}$
    for $k=1$ lump on the soliton.}
    \label{fig:1lump}
\end{figure}

Lumps in the domain wall are Skyrmions in the bulk point of view as shown below.
Such composite states of Skyrmions 
and a domain wall are called 
domain-wall Skyrmions
 \cite{Nitta:2012wi,Nitta:2012rq,
Gudnason:2014nba,Gudnason:2014hsa,
Nitta:2022ahj}\footnote{
Two-dimensional version of domain-wall skyrmions
were also earlier found in field theory 
\cite{Nitta:2012xq,Kobayashi:2013ju,Jennings:2013aea}
and studied experimentally and theoretically 
in chiral magnets 
\cite{PhysRevB.99.184412,PhysRevB.102.094402,KBRBSK,Nagase:2020imn,Yang:2021,Ross:2022vsa} (see also \cite{Kim:2017lsi}).
}. 
However, ours are crucially different from these cases;
Due to eqs.~(\ref{EFT_of_GW}) and (\ref{lump_and_Skyrme}),
the topological lump charge 
$k \in \pi_2(\mathbb{C}P^1)$ is related to the baryon number 
$N_{\rm B} \in \pi_3[SU(2)]$ (topological charge 
of Skyrmions in the bulk) by
\begin{eqnarray}
    N_{\rm B} = \int \rmd^3x\, {\cal B} = 2\int \rmd^2x\, q = 2 k.
\end{eqnarray}
Thus, $N_{\rm B}$ inside the soliton world-volume
is quantized in even integer.
We show the baryon charge density for the minimal lump $k=1$ in Fig.~\ref{fig:1lump}. It is clearly seen that the two baryons come in pairs 
in a Macarons shape
sandwiching the domain wall.

We now take into account the gauge coupling between
${\bm n}$ and $A_\alpha$. Since the $U(1)_{\rm EM}$ is generated by $\tau_3$, $n_3$ is neutral whereas $n_1 + \rmi n_2 \to \rme^{-\rmi\lambda}(n_1+\rmi n_2)$.
Thus the covariant derivative is given by $D_\alpha (n_1+\rmi n_2) = (\partial_\alpha - \rmi e A_\alpha)(n_1+\rmi n_2)$.
Let $C$ be a closed curve on which $n_3 = 0$, 
and $D$ be the interior of $C$. The $U(1)_{\rm EM}$ is 
spontaneously broken around $|n_1+\rmi n_2|=1$ 
where charged pions are condensed. 
Thus, the closed curve 
$C$ is a superconducting ring, 
and there is a persistent current along it.
Let us write $n_1+\rmi n_2 = \rme^{i\psi}$ on $C$. 
The configuration of the gauge field 
along $C$ is determined by minimizing 
the gradient energy 
$|D_\alpha (n_1+\rmi n_2)|^2 = 0$, 
yielding $\partial_\alpha \psi = e A_\alpha$.
Then, we have a flux (and area) 
quantization on $D$:
\begin{eqnarray}
B S_D= \int_D \rmd^2x\, B = \oint_C \rmd x^i A_i
= \frac{1}{e} \oint_C \rmd x^i \partial_i \psi
= \frac{2\pi k}{e} \;\; \label{quantize}
\end{eqnarray}
with the lump number $k$ on $D$  
and the area $S_D$ of $D$.
This gives a constraint among the lump moduli.

For example, a single ($k=1$) lump is given by 
$f = b_0/w$ with 
a size  and phase  moduli $|b_0|$ and arg $b_0$, 
respectively, 
and $n_3 = (|w|^2-|b_0|^2)/(|w|^2+|b|^2)$.
Thus, the size of $D$ bounded by $n_3=0$ 
is $|w|=|b_0|$, 
and the flux quantization implies
a quantization of the size $|b_0|=\sqrt{2/eB}$. 
Axially symmetric $k$-lumps 
is given by $f =b_0/w^k$ and 
$n_3 = (|w|^4-|b_0|^2)/(|w|^4+|b|^2)$.  
In this case, we have $|b_0|=\sqrt{2k/eB}$.
\begin{figure}
    \centering
    \includegraphics[width=8cm]{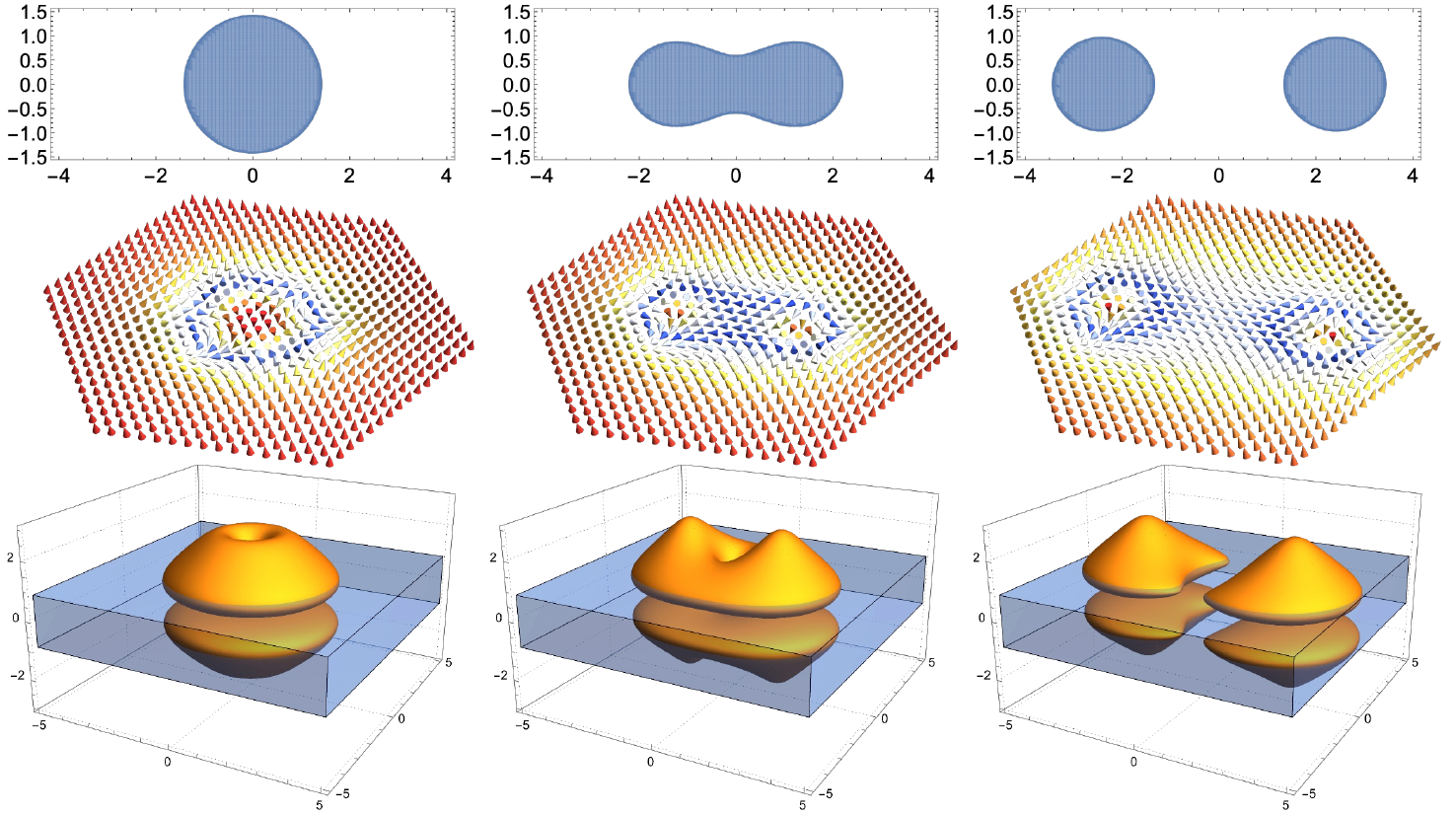}
    \caption{(top) The regions ($S_D=4\pi/eB$ with $eB=2$) in the $xy$ plane where $n_3 \le 0$ for $k=2$ lumps with $f=\frac{b_0}{(w-X)(w+X)}$ with $(b_0,X) = (2,0),\, (5,2.607),\, (2.595,1.5)$. (middle) ${\bm n}$ on the soliton. (bottom) The isosurface of ${\cal B} = 1/(15\pi^2)$.}
    \label{fig:k2_area_2pi}
\end{figure}

Finally, we discuss the condition that the lump appears 
in the ground state 
but not as an excited one.
Inserting eq.~(\ref{sol_n})
into eq.~(\ref{Bogomolny_bound}),
we get~\footnote{See Appendix 
\ref{app:WZW}
for a derivation.}
\begin{gather}
    E_{\rm DWSk}= \frac{32\pi f_{\pi}^2}{3m_{\pi}}|k|
    -2\mu_{\textrm{B}}k
    + e\mu_{\textrm{B}}B |b_{k-1}|^2\,.
    \label{E_3}
\end{gather}
The condition that the lump is energetically more 
stable than the uniform state 
$\bm{n}=(0,0,1)$, or $E<0$,
leads to the critical chemical potential for given external parameters,
$\mu_{\textrm{B}}$. 
For $k=1$, 
the flux quantization condition in eq.~(\ref{quantize}), 
or $|b_0|=\sqrt{2/eB}$, implies that 
the last two terms in eq.~(\ref{E_3}) 
cancel out, leaving the first term.  
Since this is positive, 
single lumps 
on the domain wall are always excited states, 
but not spontaneously created. 
The situation is drastically different 
for $k \ge 2$.
In this case, the energy in eq.~(\ref{quantize}) is 
minimized when 
\begin{equation}
    b_{k-1} =0, \quad 
{\rm for} \quad k \ge 2.\label{eq:b=0}
\end{equation}
This condition is consistent with the flux quantization condition eq.~(\ref{quantize}), 
unlike the case of $k=1$.
For instance, for the $k=2$ axisymmetric lump, 
$f = b_0/w^2$ due to eq.~(\ref{eq:b=0}). 
Then, the quantization condition (\ref{quantize}) imposes $|b_0| = 2/\sqrt{eB}$.
Fig.~\ref{fig:k2_area_2pi} shows the most general 
$k=2$ lump configurations 
$f=\frac{b_0}{(w-X)(w+X)}$, in which  
the $k=2$ axisymmetric lump can be separated 
into two $k=1$ lumps with keeping the total area 
and total energy.

\begin{figure}[tb]
    \centering
    \includegraphics[width=8.5cm]{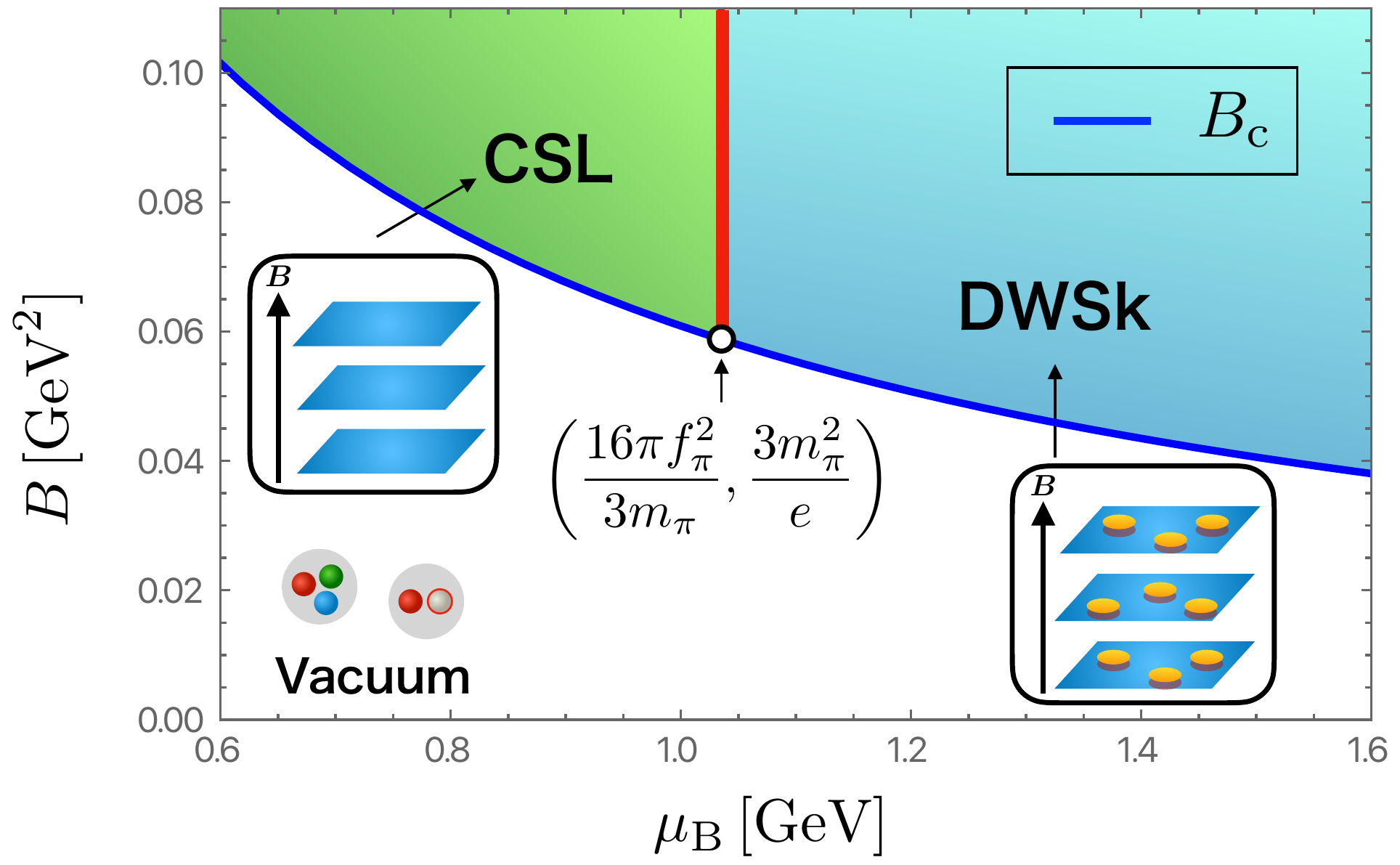}

\caption{
    The red and blue lines show $\mu_{\textrm{c}}$ and $B_{\textrm{c}}$, respectively.
    In the green or blue regions,
    the ground states are the CSL of $\pi_0$ and the domain wall skyrmion (DWSk) phase, respectively.
    }
    \label{fig:phase_diagram}
\end{figure}

The energy of the lump ($k \geq 2$) is negative 
when eq.~(\ref{eq:negative}) holds, 
yielding a transition from the CSL phase to 
the domain-wall Skyrmion phase 
in which lumps are spontaneously created.
The phase diagram <
is shown in Fig.~\ref{fig:phase_diagram}, 
in which the domain-wall Skyrmion phase meets the CSL and vacuum at 
the tricritical point 
($\mu_{\rm c},3m_{\pi}^2/e\sim 0.06 {\rm GeV}^2\sim 1.0\times 10^{19} {\rm G}$).
This  point coincides with 
the instability of CSLs via a charged pion condensation \cite{Brauner:2017uiu},
implying that the domain-wall Skyrmion phase 
is the fate of such an instability.

Apart from the binding energy between 
a Skyrmion and the domain wall, 
$\mu_{\textrm{c}}$ is nothing but 
the 
nuclear saturation density 
above which nucleons can be excited, 
or simply gives the mass of nucleons
in this media.
It is interesting to point out that 
$\mu_{\textrm{c}}$ is written thoroughly
in terms of only the pion's decay constant and mass.

\section{Summary}
We have reported the existence of 
the domain-wall Skyrmion phase 
in QCD matter at finite density with magnetic field. 
Our results were obtained at the leading 
${\cal O}(p^2)$ order of the ChPT, 
without a help of higher order corrections 
such as a Skyrme term.
Thus our results are model-independent and robust.
The lumps on the domain wall are Skyrmions(baryons) in the bulk, 
for which one lump corresponds to two baryons.
Skyrmions exist in a pair 
in a Macarons shape  
with sandwiching the domain wall. 
The boundary $n_3=0$ of the lumps are superconducting rings, 
yielding the flux quantization condition or
the area quantization of the lumps, eq.~(\ref{quantize}). 
While single $k=1$ lumps always have positive energy, 
$k(\geq 2)$ lumps have negative energy in 
the domain-wall Skyrmion phase.
As a byproduct, we have obtained 
the nuclear saturation density 
in terms of the pion's decay constant and mass.
The magnetic field $10^{19} {\rm G}$ at the tricritical point
is about only 10 times stronger than the ability of current heavy-ion colliders and so it is in scope of future experiments.

An Abrikosov's vortex lattice
was proposed \cite{Evans:2022hwr}
as a possible fate of the charged pion condensation 
\cite{Brauner:2017uiu}.
However they used ${\mathcal O}(p^4)$ order 
(but not all terms)
of the ChPT. 
In contrast, we have considered the leading  
${\mathcal O}(p^2)$ order consistently 
with all the terms.

\begin{acknowledgments}
We thank Naoki Yamamoto for useful comments.
This work is supported in part by 
 JSPS KAKENHI [Grants No. JP19K03839 (ME) and No. JP22H01221 (ME and MN)], the WPI program ``Sustainability with Knotted Chiral Meta Matter (SKCM$^2$)'' at Hiroshima University.
 K. N. is supported by JSPS KAKENHI, Grant-in-Aid for Scientific Research No. (B) 21H01084.
\end{acknowledgments}


\appendix

\section{Derivation of the effective action of a non-Abelian sine-Gordon soliton}\label{app:EFT}
Here, 
we derive the EFT of the non-Abelian sine-Gordon soliton 
for the both orientational moduli $\phi$ and the translational modulus $Z$, 
following Ref.~\cite{Eto:2015uqa} 
in the moduli approximation \cite{Manton:1981mp,Eto:2006uw}.

We first calculate the effective action coming from eq.~(\ref{ChPT_with_B}).
Substituting eq.~(\ref{general_sol_Sigma}) into eq.~(\ref{ChPT_with_B}),
we get
\begin{align}
    &\mathcal{L}_{\textrm{ChPT}}\notag\\
    &= \frac{f_{\pi}^2}{2} \left\{
    \del_{\alpha}\theta \del^{\alpha}\theta 
    + |1-u^2|^2 [(\phi^{\dag}\del_{\alpha}\phi)^2+\del_{\alpha}\phi^{\dag}\del^{\alpha}\phi]
    \right\} \notag \\ 
    &+ \frac{e}{2}f_{\pi}^2A^{\alpha}|u^2-1|^2 
    \notag\\ & \quad  \times 
    \left[
    \phi^{\dag}\tau_3\phi \cdot \phi^{\dag}\del_{\alpha}\phi
    + \frac{1}{2}(\del_{\alpha}\phi^{\dag}\tau_3\phi - \phi^{\dag}\tau_3\del_{\alpha}\phi)
    \right] \notag \\
    &- \frac{e^2 f_{\pi}^2}{8}A^2 |u^2-1|^2 \left[-1 + (\phi^{\dag}\tau_3\phi)^2 \right] \notag \\
    &- \frac{f_{\pi}^2}{2} (\del_z\theta)^2 - f_{\pi}^2m_{\pi}^2(1-\textrm{cos}\theta) \,,
\end{align}
where $x^\alpha$ ($\alpha =0,1,2$) are world-volume coordinates.
Since $\theta$ depends on $x^\alpha$ through $Z(x^{\alpha})$,
$\del_{\alpha}\theta$ becomes
\begin{gather}
    \del_{\alpha}\theta = -\del_{\alpha}Z \del_z\theta \,,
\end{gather}
then $f_{\pi}^2\del_{\alpha}\theta \del^{\alpha}\theta/2$ can be expressed as
\begin{gather}
    \frac{f_{\pi}^2}{2} (\del_z\theta)^2 \del_{\alpha}Z \del^{\alpha}Z \,.
\end{gather}
Integrating over $z$, the effective action stemming from eq.~(\ref{ChPT_with_B}) can be calculated as
\begin{align}
    &\int \rmd z\, \mathcal{L}_{\textrm{ChPT}}\notag\\
   & = 4m_{\pi}f_{\pi}^2 (\del_{\alpha}Z)^2 + \frac{16f_{\pi}^2}{3m_{\pi}} [(\phi^{\dag}\del_{\alpha}\phi)^2+\del_{\alpha}\phi^{\dag}\del^{\alpha}\phi] \notag \\
    &+ \frac{16{\rmi} e}{3m_{\pi}}f_{\pi}^2A^{\mu} \left[
    \phi^{\dag}\tau_3\phi \cdot \phi^{\dag}\del_{\mu}\phi
    + \frac{1}{2}(\del_{\mu}\phi^{\dag}\tau_3\phi - \phi^{\dag}\tau_3\del_{\mu}\phi)
    \right] \notag \\
    &- \frac{4e^2 f_{\pi}^2}{3m_{\pi}}A^2 
    \left[-1 + (\phi^{\dag}\tau_3\phi)^2 \right]
    - 8m_{\pi}f_{\pi}^2 \label{int_z_ChPT} \,,
\end{align}
where we have used the integrals,
\begin{gather}
    \int_{-\infty}^{\infty} \rmd z\, (\del_z\theta)^2 = 8m_{\pi} \,, \\
    \int_{-\infty}^{\infty} \rmd z\, |1-u^2|^2 = \frac{32}{3m_{\pi}} \,.
\end{gather}
The gauge field in the above Lagrangian can be stored
in the standard covariant derivative:
$D_{\alpha}\phi = (\del_{\alpha}+\rmi e A_{\alpha}\tau_3/2)\phi$.
This comes from the original gauge transformation
$\Sigma\to \rme^{i\lambda\frac{\tau_3}{2}} \Sigma \rme^{-i\lambda\frac{\tau_3}{2}}$ and $A_\mu \to A_\mu - \frac{1}{e}\partial_\mu \lambda$
with the definition $\Sigma = [{\bm 1}_2 + (u^2-1)\phi \phi^{\dag}]u^{-1}$ and $\phi \to \rme^{\rmi\frac{\lambda \tau_3}{2}}\phi$. Then, 
eq.~(\ref{int_z_ChPT}) is simplified into
\begin{align}
    \int \rmd z\, \mathcal{L}_{\textrm{ChPT}}
    &= -8f_{\pi}^2m_{\pi} + 4m_{\pi}f_{\pi}^2 (\del_{\alpha}Z)^2 \notag \\
    &+ \frac{16f_{\pi}^2}{3m_{\pi}} [(\phi^{\dag}D_{\alpha}\phi)^2+D_{\alpha}\phi^{\dag}D^{\alpha}\phi] \,.
\end{align}
The first term is the tension of the non-Abelian sine-Gordon soliton,
and the second and third terms are the kinetic terms of $Z$ and $\phi$, respectively.

We next calculate the effective action coming from the WZW term in eq.~(\ref{effective_Lagrangian_GW}).
Due to the eq.~(\ref{lump_and_Skyrme}),
the first term can be expressed as $\mu_{\textrm{B}}{\cal B}$.
Integration of ${\cal B}$ over $z$ equals twice $q$.
Then, we get
\begin{gather}
    \int \rmd z\, \mu_{\textrm{B}}{\cal B} = 2\mu_{\textrm{B}}q \,.
\end{gather}
The second term in eq.~(\ref{effective_Lagrangian_GW}) is divided into two terms as follows:
\begin{align}
    &\frac{\rmi e\mu_{\textrm{B}}}{16\pi^2} \epsilon^{0ijk} \del_i[A_j\tr(\tau_3L_k+\tau_3R_k)] \notag \\
    &= \frac{\rmi e\mu_{\textrm{B}}}{16\pi^2}\epsilon^{0ijk} \del_iA_j \tr(\tau_3L_k+\tau_3R_k) \notag \\
    &+ \frac{\rmi e\mu_{\textrm{B}}}{16\pi^2}\epsilon^{0ijk} A_j \tr \tau_3 (\del_i\Sigma \del_k\Sigma^{\dag} + \del_k\Sigma^{\dag}\del_i\Sigma) \label{WZW_gauged_term} \,.
\end{align}
We consider the uniform external magnetic field along the $z$-axis, $\bm{B}=(0,0,B)$.
Then, the first term in eq.~(\ref{WZW_gauged_term}) becomes
\begin{gather}
    -\frac{\rmi e\mu_{\textrm{B}}}{16\pi^2}B \tr \tau_3(L_3+R_3) \label{domain-wall_charge} \,.
\end{gather}
In terms of the projection operator 
$P\equiv \phi \phi^{\dag}$, 
$R_k$ and $L_k$ can be expressed as
\begin{gather}
    L_k = (1-2P) \rmi \del_k\theta + (u^{-2}-1)\del_kP + |u^2-1|^2P\del_kP \,, \\
    R_k = (1-2P) \rmi \del_k\theta + (u^{-2}-1)\del_kP + |u^2-1|^2\del_kP \cdot P \,.
\end{gather}
Since $\phi$ does not depend on $z$,
the second and third terms in $L_3$ and $R_3$ vanish.
Therefore, eq.~(\ref{domain-wall_charge}) becomes
\begin{gather}
    -\frac{e \mu_{\textrm{B}}B}{4\pi^2} (\phi^{\dag}\tau_3\phi) \del_3\theta \,,
\end{gather}
and integrating over $z$, we get
\begin{gather}
    \int \rmd z\,
    \frac{\rmi e\mu_{\textrm{B}}}{16\pi^2}\epsilon^{0ijk} \del_iA_j \tr(\tau_3L_k+\tau_3R_k)
    = -\frac{e \mu_{\textrm{B}}B}{2\pi} \phi^{\dag}\tau_3\phi \label{WZW_first_term} \,,
\end{gather}
where we have used the boundary condition of $\theta$ for a single soliton, $\theta(\infty)-\theta(-\infty)=2\pi$.

We next calculate the second term in eq.~(\ref{WZW_gauged_term}).
Substituting eq.~(\ref{general_sol_Sigma}) into $\del_i\Sigma \del_k\Sigma^{\dag}$ and $\del_k\Sigma^{\dag} \del_i\Sigma$, these two quantities can be calculated as
\begin{align}
    &\del_i\Sigma \del_k\Sigma^{\dag}
    = \del_i\theta \del_k\theta + [(1-u^{-2}) -(u^2 - u^{-2})P] \rmi \del_i\theta \del_kP \notag \\
    & - [(1-u^{2})+(u^2 - u^{-2})P] \rmi \del_k\theta \del_iP
    + |1-u^2|^2 \del_iP \del_kP \,,
\end{align}
and
\begin{align}
    &\del_k\Sigma^{\dag} \del_k\Sigma
    = \del_i\theta \del_k\theta + [(1-u^{-2}) -(u^2 - u^{-2})P] \rmi \del_i\theta \del_kP \notag \\
    & - [(1-u^{2})+(u^2 - u^{-2})P] \rmi \del_k\theta \del_iP
    + |1-u^2|^2 \del_kP \del_iP \,.
\end{align}
Hence, we can represent $\epsilon^{0ijk}\tr(\del_i\Sigma \del_k\Sigma^{\dag} + \del_k\Sigma^{\dag} \del_i\Sigma)$ in terms of $u$ and $\phi$ as follow :
\begin{align}
    &\epsilon^{0ijk}\tr \tau_3(\del_i\Sigma \del_k\Sigma^{\dag} + \del_k\Sigma^{\dag} \del_i\Sigma) \notag \\
    &= \epsilon^{0ijk}\tr \tau_3 \{
    2\rmi \left[-1+(1+u^2)P \right] (u^{-2}-1) \del_i\theta \del_kP \notag \\
    &\quad + 2\rmi \left[-1+(1+u^{-2})P \right] (u^{2}-1) \del_i\theta \del_kP
    \} \notag \\
    &= 2\rmi |1-u^2|^2\epsilon^{0ijk} \del_i\theta \tr \tau_3 \del_kP \notag \\
    &= 2\rmi |1-u^2|^2\epsilon^{03jk} \del_3\theta \tr \tau_3 \del_kP \,.
\end{align}
Inserting this expression into the second term in eq~(\ref{WZW_gauged_term}),
it becomes
\begin{gather}
    -\frac{e \mu_{\textrm{B}}}{8\pi^2}
    \epsilon^{03jk}|1-u^2|^2 \del_3\theta A_j\del_k(\phi^{\dag}\tau_3\phi) \,.
\end{gather}
Integrating over $z$, we get
\begin{gather}
    \int \rmd z\, \frac{\rmi e\mu_{\textrm{B}}}{16\pi^2}\epsilon^{0ijk} A_j \tr \tau_3 (\del_i\Sigma \del_k\Sigma^{\dag} + \del_k\Sigma^{\dag}\del_i\Sigma) \notag \\
    = -\frac{e \mu_{\textrm{B}}}{2\pi} \epsilon^{03jk} A_j\del_k(\phi^{\dag}\tau_3\phi) \label{WZW_second_term} \,,
\end{gather}
where we have used the integral,
\begin{gather}
    \int_{-\infty}^{\infty}\rmd z(u-u^{-1})^2\del_3\theta = -4\pi \,.
\end{gather}
Summing up eqs.~(\ref{WZW_first_term}) and (\ref{WZW_second_term}), the effective Lagrangian from the second term in eq.~(\ref{eq:GW}) can be  calculated as
\begin{gather}
    -\frac{e \mu_{\textrm{B}}}{2\pi} \epsilon^{03jk}\del_j(A_k\phi^{\dag}\tau_3\phi) \,.
\end{gather}
Finally, we arrive at the effective Lagrangian of the non-Abelian sine-Gordon soliton under the magnetic field :
\begin{align}
    \mathcal{L}_{\textrm{DW}} &\equiv \int \rmd z\, (\mathcal{L}_{\textrm{ChPT}} + \mathcal{L}_{\textrm{WZW}}) \notag \\
    &= 8f_{\pi}^2m_{\pi} - \frac{e\mu_{\textrm{B}}B}{2\pi} 
    + 4m_{\pi}f_{\pi}^2 (\del_{\alpha}Z)^2 \notag\\
    & + \frac{16f_{\pi}^2}{3m_{\pi}} [(\phi^{\dag}D_{\alpha}\phi)^2+D_{\alpha}\phi^{\dag}D^{\alpha}\phi] \notag \\
    &+ 2\mu_{\textrm{B}}q + \frac{e \mu_{\textrm{B}}}{2\pi} \epsilon^{03jk}\del_j[A_k(1-n_3)] \,,
\end{align}
which yields eqs.~(\ref{EFT_lag}), (\ref{EFT_of_kinetic}) and (\ref{EFT_of_GW}).

\section{Correspondence between 
topological charges in three and two dimensions}
\label{app:baryon}

\begin{figure}
\begin{center}
    \includegraphics[width=8cm]{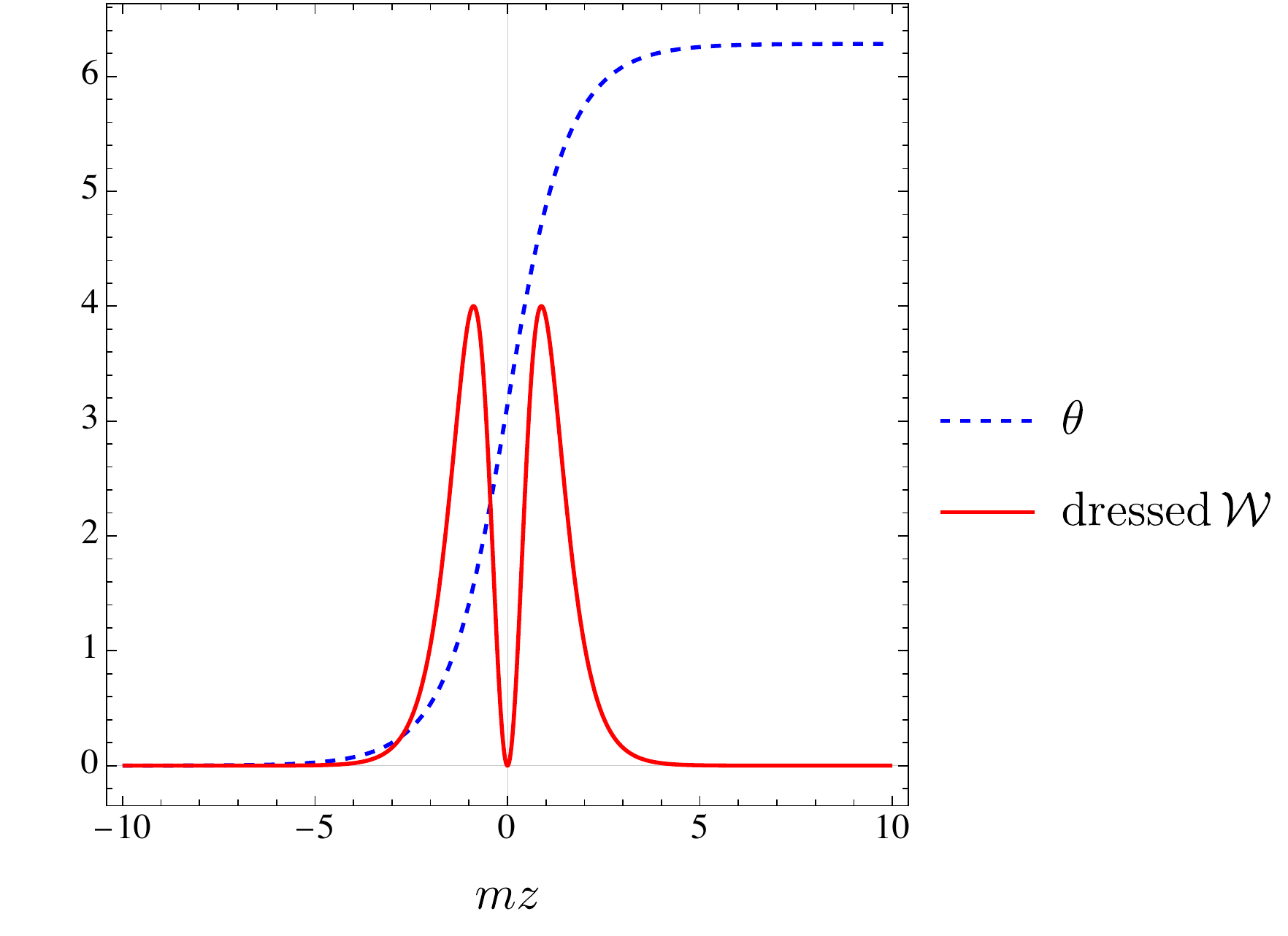}
    \caption{The single sine-Gordon soliton $\theta$ and the dressed domain wall density $-(u-u^{-1})^2 {\cal W}$.}
    \label{fig:wall_density}
\end{center}
\end{figure}

Here, we derive the correspondence between 
the topological charge in three dimensions and two dimensions 
by using the decomposition 
of the topological charge density, 
the Baryon number density, 
into the sine-Gordon soliton number density 
and the lump topological charge density.
The baryon charge density
\begin{eqnarray}
    {\cal B} = \frac{-1}{24\pi^2}\epsilon^{ijk} \tr(L_{i}L_{j}L_{k}),
\end{eqnarray}
can be factorized as
\begin{eqnarray}
    {\cal B} = -(u-u^{-1})^2 {\cal W}(z) q(x,y),
\end{eqnarray}
where ${\cal W}(w)$ is the domain wall charge density
\begin{eqnarray}
    {\cal W}(z) &\equiv& \frac{\partial_3 \theta}{2\pi} = \frac{m}{\pi \cosh mz},
\end{eqnarray}
and the lump topological charge density
\begin{eqnarray}
    q(x,y) \equiv
     -\frac{\rmi}{2\pi}\epsilon^{ij}\del_i\phi^{\dag}\del_j\phi
     = \frac{1}{8\pi} \epsilon^{ij}{\bm n}\cdot (\partial_i {\bm n}\times\partial_j{\bm n}).
\end{eqnarray}
The sine-Gordon soliton number and the lump number are given by
\begin{eqnarray}
    \int^\infty_{-\infty}dz\, {\cal W} = 1,\quad \int d^2x\, q = k,
\end{eqnarray}
respectively. 
The dressing factor in ${\cal B}$ is given by
\begin{eqnarray}
    -(u-u^{-1})^2 = \frac{16 \sinh^2mw}{\cosh^4mw}.
\end{eqnarray}
Thus, the domain wall number density together with the dressing factor
determines $z$-dependence of ${\cal B}$
\begin{eqnarray}
    -(u-u^{-1})^2 {\cal W} = \frac{16 m \sinh^2mw}{\pi\cosh^5mw}.
\end{eqnarray}
This has two peaks as shown in Fig.~\ref{fig:wall_density} and its integration reads
\begin{eqnarray}
    - \int^{\infty}_{-\infty}dz\, (u-u^{-1})^2 {\cal W} = 2.
\end{eqnarray}
Hence, the $z$ integration of the baryon number density is
\begin{eqnarray}
    \int^{\infty}_{-\infty}dz\, {\cal B} = 2 q(x,y).
\end{eqnarray}

\section{The WZW term for $k$ lumps}
\label{app:WZW}
Here, we calculate the WZW term. 
for $k$-lump configuration. 
Let us take the constant magnetic flux background with
\begin{eqnarray}
    A_i = -\frac{B}{2}\epsilon_{ij}x^j,\quad F_{12} = B.
\end{eqnarray}
Then, the WZW term includes
\begin{eqnarray}
&&
\frac{e \mu_{\textrm{B}}}{2\pi} \epsilon^{03jk}\del_j[A_k(1-n_3)]\nonumber\\
&&~= -\frac{e \mu_{\textrm{B}} B}{4\pi} 
\epsilon^{jk}\epsilon_{kl}\partial_j
\left(
x^l(1-n_3)
\right) \nonumber\\
&&~= \frac{e \mu_{\textrm{B}} B}{4\pi} 
\partial_j
\left(
x^j(1-n_3)
\right),
\end{eqnarray}
where our notation is $\epsilon^{12}=\epsilon_{12} = 1$.
Next we integrate this over the $xy$ plane
\begin{eqnarray}
    &&\int d^2x~\frac{e \mu_{\textrm{B}}}{2\pi} \epsilon^{03jk}\del_j[A_k(1-n_3)] \nonumber\\
    &&~=
    \frac{e \mu_{\textrm{B}} B}{4\pi} 
    \int d^2x~
    \partial_j
    \left(
    x^j(1-n_3)
    \right) \nonumber\\
    &&~= \frac{e \mu_{\textrm{B}} B}{4\pi} \oint dS_j x^j(n_3-1).
\end{eqnarray}
The lump solution can be easily described by 
the inhomogeneous coordinate $f \in \mathbb{C}$ by
\begin{eqnarray}
    \phi = \frac{1}{\sqrt{1+|f|^2}}\left(
    \begin{array}{c}
    1\\
    f
    \end{array}
    \right),
\end{eqnarray}
providing
\begin{eqnarray}
    \phi\phi^\dag 
    = \frac{1}{1+|f|^2}
    \left(
    \begin{array}{cc}
    1 & \bar f\\
    f & |f|^2
    \end{array}
    \right).
\end{eqnarray}
Then, the equation of motion reduces to
\begin{eqnarray}
\partial_{\bar{w}} f = 0,
\end{eqnarray}
with $w \equiv x + i y$.
Hence $f$ can be an arbitrary meromorphic function
\begin{eqnarray}
    f(w) = \frac{b_{k-1}w^{k-1}+\cdots+b_1w+b_0}{w^k + a_{k-1}w^{k-1}+\cdots+a_1w+a_0}.
\end{eqnarray}
The highest power $k$ is the lump charge.
The $n_3$ component for the $k$ lumps is given by
\begin{eqnarray}
    n_1 &=& \phi^\dag \tau_1 \phi = \frac{f+\bar f}{1+|f|^2}\,,\\
    n_2 &=& \phi^\dag \tau_2 \phi = \frac{-i(f-\bar f)}{1+|f|^2}\,,\\
    n_3 &=& \phi^\dag \tau_3 \phi = \frac{1-|f|^2}{1 + |f|^2}.
\end{eqnarray}
We have $n_3 = 1$ for zeros of $f$,
whereas $n_3 = -1$ for poles of $f$.
In order to evaluate the WZW term, it is enough
for us to know the asymptotic behavior
\begin{eqnarray}
    n_3 -1 
    \to \frac{r^2-|b_{k-1}|^2}{r^2+|b_{k-1}|^2} - 1
    = \frac{-2|b_{k-1}|^2}{r^2+|b_{k-1}|^2}.
\end{eqnarray}
Hence, we get
\begin{eqnarray}
    &&\frac{e \mu_{\textrm{B}} B}{4\pi} \oint dS_j x^j(n_3-1)\nonumber\\
    &&~= \frac{e \mu_{\textrm{B}} B}{4\pi} \lim_{r\to\infty}\int^{2\pi}_0d\theta~ r^2\frac{-2|b_{k-1}|^2}{r^2+|b_{k-1}|^2} \nonumber\\
    &&~= - e \mu_{\textrm{B}} B|b_{k-1}|^2.
\end{eqnarray}
Thus, this term contributes to the energy
\begin{eqnarray}
    E \supset e \mu_{\textrm{B}} B|b_{k-1}|^2.
\end{eqnarray}
This implies that $b_{k-1} = 0$ is energetically favored.


\providecommand{\noopsort}[1]{}\providecommand{\singleletter}[1]{#1}%

\end{document}